\documentclass[12pt]{iopart}
\usepackage{graphicx}
\usepackage{bm}
\usepackage{iopams}
\usepackage{color}
\usepackage{amssymb}
\usepackage[caption = false]{subfig}
\newcommand{\mainmatter}{%
  \setcounter{footnote}{0}%
  \patchcmd{\@makefntext}{\fnsymbol}{\arabic}{}{}%
  \patchcmd{\@thefnmark}{\fnsymbol}{\arabic}{}{}%
  \def\@makefnmark{\textsuperscript{\arabic{footnote}}}%
}
\newcommand{\beqar}{\begin{eqnarray}}
\newcommand{\eeqar}{\end{eqnarray}}
\newcommand{\bea}{\begin{eqnarray}}
\newcommand{\eea}{\end{eqnarray}}
\newcommand{\bcen}{\begin{center}}
\newcommand{\ecen}{\end{center}}

\newcommand{\bra}[1]{\left< #1 \right|}
\newcommand{\ket}[1]{\left| #1 \right>}

\newcommand{\f}[2]{\frac{#1}{#2}}
\renewcommand{\b}[1]{\left({#1}\right)}
\renewcommand{\v}[1]{\vec{#1}}
\newcommand{\pd}[2]{\frac {\partial #1}{\partial #2}}

\renewcommand{\sb}[1]{\left[{#1}\right]}
\newcommand{\mean}[1]{\langle {#1} \rangle}

\newcommand{\ra}{\rightarrow}

\renewcommand{\b}[1]{\left({#1}\right)}
\renewcommand{\v}[1]{\vec{#1}}

\newcommand{\eps}{\varepsilon}
\newcommand{\lam}{\lambda}

\begin{document}
\title{Experimental verification of the inertial theorem control protocols}

\author{Chang-Kang Hu}
\footnote[1]{Equal contribution of Chang-Kang Hu and Roie Dann}
\address{CAS Key Laboratory of Quantum Information, University of Science and Technology of China, Hefei 230026, People's Republic of China}
\address{CAS Center For Excellence in Quantum Information and Quantum Physics, University of Science and Technology of China, Hefei 230026, People's Republic of China}

\author{Roie Dann}
\footnote[2]{}
\ead{roie.dann@mail.huji.ac.il}
\address{The Institute of Chemistry, The Hebrew University of Jerusalem, Jerusalem 9190401, Israel}
\address{Kavli Institute for Theoretical Physics, University of California, Santa Barbara, CA 93106, USA}

\author{Jin-Ming Cui}
\ead{jmcui@ustc.edu.cn}
\address{CAS Key Laboratory of Quantum Information, University of Science and Technology of China, Hefei 230026, People's Republic of China}
\address{CAS Center For Excellence in Quantum Information and Quantum Physics, University of Science and Technology of China, Hefei 230026, People's Republic of China}

\author{Yun-Feng Huang}
\ead{hyf@ustc.edu.cn}
\address{CAS Key Laboratory of Quantum Information, University of Science and Technology of China, Hefei 230026, People's Republic of China}
\address{CAS Center For Excellence in Quantum Information and Quantum Physics, University of Science and Technology of China, Hefei 230026, People's Republic of China}

\author{Chuan-Feng Li}
\ead{cfli@ustc.edu.cn}
\address{CAS Key Laboratory of Quantum Information, University of Science and Technology of China, Hefei 230026, People's Republic of China}
\address{CAS Center For Excellence in Quantum Information and Quantum Physics, University of Science and Technology of China, Hefei 230026, People's Republic of China}

\author{Guang-Can Guo}
\address{CAS Key Laboratory of Quantum Information, University of Science and Technology of China, Hefei 230026, People's Republic of China}
\address{CAS Center For Excellence in Quantum Information and Quantum Physics, University of Science and Technology of China, Hefei 230026, People's Republic of China}

\author{Alan C. Santos}
\ead{ac\_santos@df.ufscar.br}
\address{Instituto de F\'{i}sica, Universidade Federal Fluminense, Av. Gal. Milton Tavares de Souza s/n, Gragoat\'{a}, 24210-346 Niter\'{o}i, Rio de Janeiro, Brazil}
\address{Departamento de Física, Universidade Federal de S\~{a}o Carlos, Rodovia Washington Luís, km 235 - SP-310, 13565-905 S\~{a}o Carlos, SP, Brazil}

\author{Ronnie Kosloff}%
 \ead{ronnie@fh.huji.ac.il}
\address{The Institute of Chemistry, The Hebrew University of Jerusalem, Jerusalem 9190401, Israel}
\address{Kavli Institute for Theoretical Physics, University of California, Santa Barbara, CA 93106, USA}

\date{\today}

\begin{abstract}
An experiment based on  a trapped Ytterbium ion validates the inertial theorem for the $SU(2)$ algebra. The qubit is encoded within the hyperfine states of the atom and controlled by RF fields. The inertial theorem generates an analytical solution for non-adiabatically driven systems that are `accelerated' slowly, bridging the gap between the sudden and adiabatic limits.  
These solutions are shown to be stable to small deviations, both experimentally and theoretically. As a result, the inertial solutions pave the way to rapid quantum control of closed, as well as open quantum systems.  For large deviations from the inertial condition, the amplitude diverges while the phase remains accurate.
\end{abstract}

%
%
%
%
\break



\section{Introduction}
Progress in contemporary quantum technology requires  precise control of quantum dynamics
\cite{glaser2015training,cirac1995quantum,monroe1995demonstration,barreiro2011open,rosi2013fast,mandel2003coherent,bloch2008many,jaksch2000fast,duan2001geometric,jonathan2000fast,nielsen2002quantum,loss1998quantum,kadowaki1998quantum,finnila1994quantum,brooke1999quantum,venegas2018cross,johnson2011quantum,santoro2002theory,rossnagel2016single,pekola2015towards}.
To answer the demand, a "universal" vocabulary of control techniques has emerged. They have been applied across a broad range of  experimental platforms,
such as NV-centers \cite{doherty2013nitrogen,doherty2012theory,bar2013solid}, trapped ions  \cite{cirac1995quantum,haffner2008quantum,kielpinski2002architecture}, and Josephson devices \cite{makhlin2001quantum,martinis2002rabi,svetitsky2014hidden}.  These techniques are encapsulated within the theoretical framework of quantum control theory \cite{koch2016controlling,d2007introduction,glaser2015training,brif2010control}. 

This theory formulates the control problem by addressing three main topics:
\begin{enumerate}
\item{Controllability, i.e., the conditions on the dynamics
that allow obtaining the objective.}
\item{Constructive mechanisms of control, the problem of
synthesis.}
\item{ Optimal control strategies and quantum speed limits.}
\end{enumerate}
The first issue controlability of unitary  dynamics of  closed quantum system has been formulated employing  Lie algebra techniques \cite{d2007introduction,huang1983controllability,jurdjevic1972control}. In this case,
the Hamiltonian of the system is separated into drift and control terms
\begin{equation}
\hat H\b t = \hat H_0 +\sum_j g_j\b t  \hat G_j~~,
\label{eq:chamil}
\end{equation}
where $\hat H_0$ is the free system Hamiltonian, $g_j(t)$ are the control fields and $\hat G_j$ are control operators.
The system is unitary controllable provided that the Lie algebra, 
spanned by the nested commutators of $\hat H_0 $ and $\hat G_j$, is full rank \cite{d2007introduction,huang1983controllability,jurdjevic1972control,ramakrishna1996relation}. 

When addressing the quantum control challenge, it is reassuring that a solution exists, nevertheless, the practical problem of finding a control protocol has not been solved. For this task
a pragmatic approach has been developed, formulating the control problem as an optimization problem, leading to optimal control theory \cite{kosloff1989wavepacket,zhu1998rapid,palao2002quantum,glaser2015training}. This approach has achieved significant success in solving specific control problems. However, the drawback is that obtaining the control protocol
relies on a specific numerical scheme which might be difficult to obtain and to generalize \cite{machnes2011comparing}.

The present study is devoted to the experimental study of constructive mechanisms of control. 
Experimental realization based on quantum control impose additional requirements: (i)The control protocol should be robust
under experimental errors and  (ii) the mechanism should be clear and simple to generalize. 
These considerations have singled out the adiabatic protocols which have  dominated the control field, across all platforms \cite{vitanov2017stimulated,albash2018adiabatic}.
Adiabatic methods are based on the adiabatic theorem
which loosely states that the system will follow an eigenvalue of the instantaneous Hamiltonian, provided that the change  in time is slow relative to the time associated with the relevant energy gaps \cite{born1928beweis,comparat2009general}.
The fact that the Hamiltonian is an invariant of the dynamics enables  a simple implementation of the control protocol by choosing the initial and final states as eigenstates of the Hamiltonian. 
The adiabatic condition on the change in the Hamiltonian will then
generate the desired transition. The
robustness of such a protocol stems from the redundancy in the intermediate Hamiltonian,
which allow variations in the protocol, provided 
the changes are sufficiently slow. This  implies that the adiabatic protocol timescale is large relative to the system free dynamics. 
The relatively long protocol durations mean that the adiabatic protocols become prone to environmental noise. 
This  fact is one of the major disadvantages of the adiabatic method.

The present study is devoted to an experimental exploration for rapid alternative control protocol, which are
based on the inertial theorem \cite{dann2021inertial}. Such protocols are termed inertial protocols and are based on
time-dependent invariants of the dynamics, beyond the adiabatic approximation. They serves as natural replacements of the instantaneous Hamiltonian of the adiabatic protocols. 
The inertial theorem follow a similar procedure as the standard adiabatic theorem \cite{schiff1968quantum}. As a consequence, the inertial and adiabatic solutions share a similar structure, which
implies that the positive features of robustness and simplicity 
are maintained without paying the price of long timescales. The experimental demonstration of the theory is based on the $SU(2)$ algebra, which is realized by $^{171}$Yb$^+$  ion confined in a Paul trap \cite{brown1991quantum}

\section{Inertial theory and solution}

For a quantum control scheme to be generic, it has to rely on simple principles that apply across many platforms. 
The control procedure requires the formulation of a dynamical map $\Lambda_t$ from an initial state $\hat{\rho}\b 0$, to the final state 
$\hat \rho (t) = \Lambda_t \hat \rho (0) = \hat U \hat \rho(0) \hat U^{\dagger} $.
The dynamical map is generated by the control Hamiltonian Eq. (\ref{eq:chamil}):
\begin{equation}
i  \frac{\partial}{\partial t} \hat U(t) = \hat H(t) \hat U(t)~~~~\textrm{with} ~~~~\hat U(0) =\hat I~~,
    \label{eq:map}
\end{equation}
where the convention $\hbar=1$ is used throughout this paper.

The major obstacle  in generating such a map  from a time-dependent control Hamiltonian is the time-ordering operation,
resulting from the fact that $[\hat H(t),\hat H(t')] \ne 0$.  The adiabatic control circumvents this problem by employing
a slow drive $g_j(t)$, allowing an approximate description in terms of the instantaneous eigenstates \cite{messiah2003quantum,comparat2009general,mostafazadeh1997quantum,sarandy2005adiabatic,kato1950adiabatic}. 
At the other extreme, the sudden limit,  
the control is so fast that 
it overshadows the dynamics generated by the drift Hamiltonian $\hat H_0$. This leads to an instantaneous change of the Hamiltonian, while leaving the system's state unaffected. 

The inertial dynamics and control paradigm serves as a compromise between the two extremes.
It is based on the inertial theorem \cite{dann2021inertial}, which introduces an explicit solution of the dynamical map $\Lambda_t$ under certain restrictions.
The  theorem is formulated in Liouville space, a vector space of system operators $\{\hat{X} \}$, endowed with an inner product $\b{\hat{X_i},\hat{X_j}}\equiv\textrm{tr}\b{\hat{X_i}^{\dagger}\hat{X_j}}$ \cite{fano1957description,von2018mathematical,gilmore2012lie}. 
In Liouville space, the system's dynamics are represented in terms of a basis of orthogonal operators  $\{\hat{B}\}$, spanning the space.  
For example, the currently studied $SU(2)$ algebra  
can be completely characterized by a time-independent operator basis
constructed from the Pauli operators
$\{ \hat I,\hat \sigma_x, \hat \sigma_y,\hat \sigma_z \}$. The chosen (ordered) operator basis then defines a state in Liouville space. Note, that a time-dependent operator basis can also be chosen, $\{\v{v}\b t\}\equiv\{\hat{v}_1\b t,\dots, \hat{v}_N\b t\}^T$, where $N$ the Liouville space dimension. This possibility serves as a major component in the inertial theorem and construction of inertial solutions.  

The dynamics in Liouville space can be solved by substituting the chosen basis $\v v\b t$ into the Heisenberg equation of motion,
\begin{equation}
 \f{d}{dt}\v{v}^H \b t = \hat{U}^\dagger\b{t,0}\sb{\b{i \sb{\hat{H}\b t,\bullet} +\pd{}{t}}\v{v} \b t}\hat{U}^\dagger\b{t,0} ~~,
 \label{dynamics Liouv}
 \end{equation}
 where superscript $H$ signifies that the operators are in the Heisenberg picture.
 
We next consider a finite time-dependent basis, forming a closed Lie algebra, this guarantees that  Eq. (\ref{dynamics Liouv}) can be solved within the basis \cite{alhassid1978connection}. For a closed Lie algebra, equation (\ref{dynamics Liouv}) has the simple form
\begin{equation}
     \f{d}{dt}\v v^H \b t = {-i {\cal{M}}\b t} \v v^H \b t~~,
     \label{Schro Liouv1}
\end{equation}     
where ${\cal{M}}\b t$ is a finite  matrix with time-dependent elements and  $\v v\b t$ is a vector \footnote{For the case of compact Lie algebras and unitary dynamics,  $\cal{M}$ is guaranteed to be Hermitian.}.


The inertial solutions are obtained by searching for a driving protocol that allows solving Eq. (\ref{Schro Liouv1}) explicitly. These then enable extending the exact solutions for a broad range of protocols employing the inertial approximation.
By choosing a unique driving protocol and the suitable time-dependent operator basis, the dynamical equation  can be expressed as
\begin{equation}
 {\cal{M}}\b t = {\cal P}\b{\v \chi}{\cal D}\b{\v{\chi},\v{\Omega}}{\cal P}^{-1}\b{\v \chi}~~.
 \label{eq:factorized}
\end{equation}
Here, ${\cal{P}}\b{\v{\chi}}$ is an invertible matrix, which depends on the inertial coefficients $\{\chi_k\}$ (for conciseness they are expressed in terms of the vector $\v{\chi}=\{\chi_1,\dots,\chi_K\}$), and
${\cal{D}}= \textrm{diag}\b{\lam_1\b{\v \chi}\Omega_1\b t,...,\lam_N\b{\v \chi}\Omega_N\b t}$ is a diagonal matrix, whose elements depend on time-dependent frequencies $\v{\Omega}\b t=\{\Omega_1\b t,\dots,\Omega_N\b t\}$, and coefficients $\{\lam_k\}$.
Such a time-dependent operator basis always exists, however, finding an analytical solution may be difficult and requires ingenuity, see \cite{dann2021inertial} Sec. V and \cite{dann2020thermodynamically} Sec. VIII for further details.

Substituting the general decomposition Eq. (\ref{eq:factorized}) into the dynamical equation, Eq. (\ref{Schro Liouv1}) leads to an exact solution for $\v v^H\b t$
\begin{equation}
 \v v^H \b{t} = \sum_{k=1}^{N^2}c_k\v{F_k}\b{\v{\chi}} e^{-i \lam_k \theta_k\b t}~~,
 \label{constat v}
\end{equation}
where the scaled-time parameters are $\theta_k\b t=\int_0^t dt' \Omega_k\b{t'}$ and $c_k=\sum_{i}{\cal P}_{ik}$ are constant coefficients. The Liouville vector  $\v {F_k}$ corresponds to the eigenoperator $\hat{F}_k=\sum_{i}{\cal{P}}^{-1}_{ki} \hat{V}_i$, where ${\cal{P}}^{-1}_{ik}$ are elements of ${\cal{P}}^{-1}$. For a Hermitian $\cal M$, the eignvalues $\lambda_k$ are either zero or are pairs with equal magnitude and opposite signs.

The solution (\ref{constat v}) is exact, but is limited to protocols for which $\v \chi$ is constant.  This serves as a very severe constraint on the possible control protocols. However, the restriction can be loosened by utilizing the inertial theorem, which introduces approximate solutions for protocols with slowly varying $\v\chi \b t$. 

For a state $\v{v}$, driven by a inertial protocol, the system's evolution is given by
\begin{eqnarray}
 \v v^H\b{t}=\sum_{k=1}^{N^2} c_k\b{\v{\chi}\b t} e^{-i\int_{0 }^{t}dt'\lam_{k}\Omega_k}e^{i \phi_{k}\b t}\v{F}_{k}\b{\v \chi\b{t}}\\
  ={\cal{P}}\b{\v \chi \b t}e^{-i\int_{\theta_k\b 0}^{\theta_k \b t} {\lambda_k\b{\theta'_k}d\theta'_k}}
    {\cal{P}}^{-1}\b{\v \chi \b t}\v{v}^H\b 0~~,
    \nonumber
 \label{eq:inetrial state}
\end{eqnarray}
where the first exponent is determined by the dynamical phase and the second includes a new geometric phase
\begin{equation}
\phi_k\b{t}= i\int_{\v{\chi} \b{ 0}}^{\v{\chi} \b{t}} d\v{\chi}\b{\v{G}_{k},\nabla_{\v{\chi}} \v F_{k}} ~~. 
\label{eq:geometric}
\end{equation}
Here, $\v{G}_{k}$ are the bi-orthogonal partners of $\v F_{k}$. 
The inertial solution is characterized by two timescales: the fast timescale is incorporated within the frequencies $\Omega_k(t)$, while 
the slow timescale is associated with the change in the inertial parameters $\chi_k \b t$.

The system's state follows the instantaneous solution determined by the instantaneous $\v{\chi} \b{t}$ and phases, associated with the eigenvalues $\lam_k\Omega_k$ and eigenoperators $\hat F_k$. We restrict the analysis to the case where $\lam_k\Omega_k$ do not cross, hence, the spectrum of ${\cal{D}}$ remains non-degenerate throughout the evolution.
 Substituting the inertial solution, Eq. (\ref{eq:inetrial state}), into Eq. (\ref{Schro Liouv1}) enables assessing the  validity of the approximation in terms of the `inertial parameter'
 \begin{equation}
 \Upsilon = \sum_{n,k}\Bigg|\f{\b{\v{G}_{k},\nabla_{\v{\chi}}{\cal M}\v F_{n}}}{\b{\lam_{n}\Omega_n-\lam_{k}\Omega_k}^{2}}\b{\f{d\v\chi}{dt}}^2\Bigg|~~.
 \label{eq:Upsilon}
 \end{equation}
 This implies that the inertial solution, Eq. (\ref{eq:inetrial state}), remains valid when $\v{\chi}$ follows a path in the parameter space of $\{ \chi_k\}$, where the eigenvalues $\lambda_k$ and $\lambda_n$ are distinct \cite{kato1950adiabatic}. 
 
Overall, the inertial solution is a linear combination of the instanteneous eigenoperators $\{\hat{F}_k\}$, and holds for slow variation of $\v \chi$, i.e., $d\v \chi/dt\ll1$, $\Upsilon\ll 1$. Physically, the condition on $d\v \chi/dt$, is associated with a slow `adiabatic acceleration' of the driving \cite{dann2021inertial}. In the adiabatic limit, decomposition Eq. (\ref{eq:factorized}) is satisfied instantaneously, where $\v{\chi}\ll1$,   and the inertial solution converges to the adiabatic result.

\subsection{Inertial solution for an SU(2) algebra}

We will demonstrate the inertial solution in the context of the $SU(2)$ algebra. The simplest realization is
by a Two-Level-System (TLS).
For the demonstration, we choose a dynamical map $\Lambda_t$ that varies the energy scale and controls the relation between energy and coherence in a non-periodic fashion.
The control Hamiltonian is chosen as:
\begin{equation}
\hat{H}\b t =  \f{1}{2}\b{\omega\b t  \hat \sigma_{z} + \varepsilon\b t \hat \sigma_{x}}~~,  
\label{eq:Ham}
\end{equation}
where the control protocol are parameterized as follows
\begin{eqnarray}
\begin{array}{lcl}
\omega\b t &=& \Omega\b t\cos\b{\alpha\b{t}t}\\ 
\varepsilon \b t &=& \Omega\b t \sin\b{\alpha\b{t}t}
\label{eq:protocol}
\end{array}
~~.
\label{eq:omega eps}
\end{eqnarray}
Here, the frequencies $\omega$ and $\epsilon$ are the detuning and Rabi frequency, respectively. These define the generalized Rabi frequency $\Omega \b t\equiv \sqrt{\epsilon^{2}\b t+\omega^{2}\b t}$.

We choose a  time-dependent operator basis 
which can factorize the equation of motion
$\v {v }^H\b t=\{ \hat{H}\b t,\hat{L}\b t,\hat{C}\b t,\hat{I}\}^T$, 
where
\begin{eqnarray}
\nonumber
\hat{L}\b t=\b{\epsilon\b t\hat{\sigma}_{z}-\omega\b t\hat{\sigma}_{x}}/2\\ 
\hat{C}\b t=\b{\Omega \b t/2}  \hat{\sigma}_{y} ~~,
\end{eqnarray}
and $\hat{I} $ is the identity operator. 

Since $\hat I$ is a constant of motion,
a reduction to a $3 \times 3$ vector space in the basis $\{\hat{H}\b t,\hat{L}\b t,\hat{C}\b t\}$ is sufficient for the dynamical description.
Following the general procedure, we calculate the dynamics of $\v{ v}^H \b t$, Eq. (\ref{dynamics Liouv}), to obtain a generator of the form
\begin{equation}
{\cal M}_{TLS}\b t = \Omega\b t {\cal B \b{\mu}}~~.   
\label{eq:factorized_TLS}
\end{equation}
with
 \begin{equation}
   {\cal{B}}\b{\mu}\equiv i\f{\dot{\Omega}}{\Omega^{2}}{\cal{I}}+{\cal{B}}'\b{\mu}~~,
\label{eq:Bmodel1}
 \end{equation}
 and
  \begin{equation}
   {\cal{B}}'\b{\mu}\equiv i\sb{\begin{array}{ccc}
0 & \mu & 0\\
-\mu & 0 & 1\\
0 & -1 & 0
\end{array}}~~.
\label{eq:Bmodel}
 \end{equation}
We can now identify the inertial coefficient $\v \chi =\chi=\mu$ with the adiabatic parameter of Hamiltonian, Eq. (\ref{eq:Ham}), it is defined as  
\begin{equation}
    |\mu\b t| \equiv \f{\dot{\omega}\epsilon-\dot{\epsilon}\omega}{\Omega^3}\sim\sum_{n\neq m} \frac{|\bra{E_{m}\b t}\dot{\hat{H}}\b t\ket{E_{n}\b t}|}{\b{E_{m}\b t - E_{n}\b t}^2} ~~.
    \label{eq:mu}
\end{equation}
Defining the scaled time $\theta\b t=\int_0^t{\Omega\b{t'}dt'}$ and decomposing the system state as
\begin{equation}
\v {v}^H \b t = \v u^H\b t \exp{\int_0^t{\f{\dot{\Omega}}{\Omega}dt'}}=\f{\Omega\b t}{\Omega\b 0} \v u^H\b t
\label{ap:v u eq}
\end{equation}
leads to a time-independent equation for $\v u^H\b \theta$    
\begin{equation}
    \f d{d\theta}\v u^H\b \theta=\sb{\begin{array}{ccc}
0 & \mu & 0\\
-\mu & 0 & 1\\
0 & -1 & 0
\end{array}}\v u^H\b{\theta}~~~.
\label{eq:u_dynm}
\end{equation}
For a constant adiabatic parameter $\mu$, we solve Eq. (\ref{eq:u_dynm}) by diagaonalization and obtain a solution in terms of the basis of eigenoperators $\v F =\{\hat{F}_1,\hat{F}_2,\hat{F}_3,\hat{I}\}^T$. The solution reads
 \begin{equation}
      \v F\b t=e^{-i{\cal{D}}\theta\b t}\v F\b{ 0 }~~,
 \label{eq:F}
 \end{equation}
where ${\cal{D}}=\textrm{diag}\b{0,\kappa,-\kappa,0}$ with $\kappa = \sqrt{1+\mu^2}$. 
The eigenoperators $\hat{F}_k$ are associated with the eigenvectors of ${\cal{B}}'$. 
The eigenoperators are calculated with the help of the diagonalization matrix $\cal{P}$: $\v{F}_i=\sum_j {\cal{P}}^{-1}_{ij}\v{u}_j$.  In the $\v {v}\b t=\{\hat{H}\b t,\hat{L}\b t,\hat{C}\b t,\hat{I} \}$ basis the eigenoperators can be written as:
\begin{eqnarray}
\begin{array}{l}
\v F_1=\f{\mu}{\kappa^{2}}\{1,0,\mu,0\}^{T} \\
\v F_2=\f 1{2\kappa^{2}}\{-\mu,-i\kappa,1,0\}^{T}\\   
\v F_3 =\f 1{2\kappa^{2}}\{-\mu,i\kappa,1,0\}^{T}~~,
\label{f-vec}
\end{array}
\end{eqnarray}
with corresponding eigenvalues are $\lam_1=0$ ,   $\lam_2={\kappa}$ and  $\lam_3 =-{\kappa}$. 
The vector $\v F_1$ corresponds to a time dependent constant of motion i.e. $\mean{ \hat F_1\b t} =\textrm{ const}$, with $\hat{F}_1\b t=\f{\mu}{\kappa^{2}}\b{\hat{H}\b t+\mu\hat{C}\b t}$.
Any system observable can be expressed in terms of the eigenoperators $\hat{F}_k$, at initial time, and the exact evolution is then given by equation (\ref{eq:F}).  

The exact solution relied on the condition of a constant adiabatic parameter, leading to the factorization Eq. (\ref{eq:factorized}). Such factorization enables employing the inertial theorem to extend the exact solution for a slow change in the adiabatic parameter ($\dot{\mu}\ll1$), leading to an analogous equation to   Eq. (\ref{eq:inetrial state}).
Making use of Eq. (\ref{ap:v u eq}) and the definition of $\v{F}_k$ Eq. (\ref{f-vec}), the solution of the $SU(2)$ dynamics becomes (the geometric phase vanishes in this case)
\begin{equation}
        \v {v}^H\b{ t}=\f{\Omega\b t}{\Omega\b 0}{\cal{P}}\b{\mu\b t}e^{-i\int_{0}^{t}{\cal{D}}\b{\mu\b{t'}}\Omega\b{t'}dt'}
    \times{\cal{P}}^{-1}\b{\mu\b t}\v {v}^H\b{ 0}~~.
    \label{eq:inertial_TLS}
    \end{equation}

\begin{figure}[t!]
\centering
\includegraphics[width=0.35\textwidth]{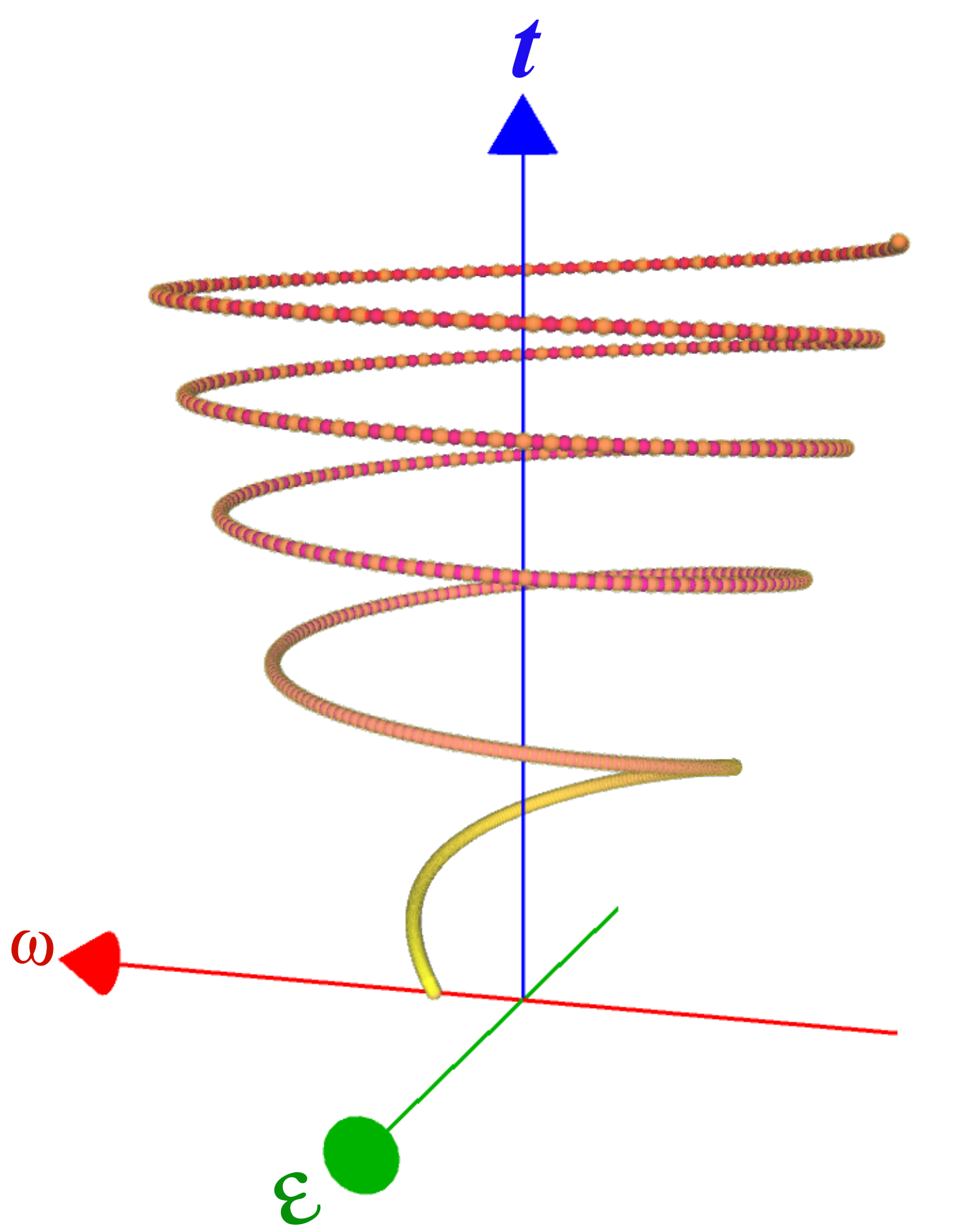}
\caption{Typical control field $\omega\b t$ and $\eps\b t$ as a function of time. Parameters correspond to the protocol of Fig. \ref{fig:1} Panel (d).  Notice the change in frequency (chirp) and change in the generalized Rabi frequency $\Omega\b t$.} 
\label{fig:control}
\end{figure}

We experimentally verify the inertial solution
by choosing a protocol associated with a linear change in the adiabatic parameter
so that $\frac{d \mu}{dt}= \delta$
\begin{equation}
    \mu\b t= \mu\b 0 + \delta \cdot t~~
    \label{eq:lin mu}
\end{equation}
and consider a linear chirp of the protocol frequencies
\begin{equation}
    \alpha\b t=\alpha\b 0 +\gamma\cdot t~~.
    \label{eq:alpha}
\end{equation}
Equations (\ref{eq:lin mu}) and (\ref{eq:alpha})  determine the Rabi frequency, by substituting this relation into Eq. (\ref{eq:mu}) we obtain $\Omega\b{t} =  -\f{\alpha\b 0+2\dot{\alpha}\b t t}{\mu}$. For this protocol, the frequencies $\omega \b t$ and $\epsilon \b t$ become
\begin{eqnarray}
\begin{array}{lcl}
 \omega\b t &=&-\f{\b{\alpha\b 0+2\gamma\cdot t}}{\mu\b 0 +\delta \cdot t}\cdot \cos\b{\b{\alpha\b 0 +\gamma t}\cdot t}\\
      \epsilon\b t &=& -\f{\b{\alpha\b 0+2\gamma \cdot t}}{\mu\b 0 +\delta \cdot t}\cdot \sin\b{\b{\alpha\b 0 +\gamma t}\cdot t}
      \end{array}
      ~~.
      \label{eq:protocol_exp}
\end{eqnarray}
A typical control field  corresponding to the frequencies $\omega\b t$ and $\eps\b t$ is shown in Fig. \ref{fig:control}, showing an evident change in frequency and amplitude.

The quality of the inertial approximation is directly connected to the parameter $\delta$. For small $|\delta|$, the inertial approximation is satisfied and the inertial solution remains accurate. 
The accuracy of the inertial solution can be evaluated by
utilizing the time-dependent control protocol, Eq. (\ref{eq:protocol_exp}). We choose the initial condition $\v{v}\b{0}=\{\hat{H}\b{0},0,0,1 \}$ which describes the system in the ground state ($\mean{\hat{H}\b 0}=-\Omega\b{0}/2$).
For these conditions, we compare the experimentally measured normalized energy, $\mean{\hat{H}\b t}/\mean{\hat{H}\b 0}$, to the inertial solution, Eq. (\ref{eq:inertial_TLS}), and a converged numerical calculation of Eq. (\ref{Schro Liouv1}), which is generated by the Hamiltonian Eq. (\ref{eq:Ham}). 

\section{Experimental setup}

\begin{figure}[t!]
	\centering
	\includegraphics[scale=1.75]{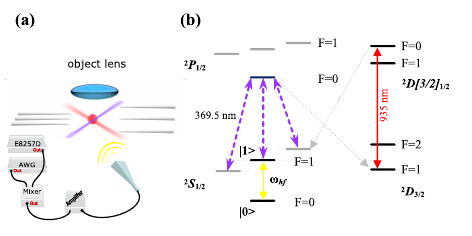}
	\caption{Experimental apparatus (a) and relevant Ytterbium energy levels (b) used in the experiment. The yellow color (waves and arrows) designates the RF transition, while the purple color signifies the doppler cooling laser transition. The grey arrow represent the spontaneous emission for the $P$ and $D$ manifolds, and the red color designates an additional optical transition, employed in order to close the cooling cycle. The qubit is encoded in the hyperfine states of the $^{2}S_{1/2}$. The readout is performed by fluorescence detection, utilizing the $395.5$ nm transition. 
	where we highlight the encoding of the two-level system used in our experimental implementation.}
	\label{ExpSche}
\end{figure}

The experimental analysis of the inertial solution employs a single Ytterbium ion $^{171}$Yb$^+$,  trapped in the six needles Paul trap, schematically shown in Fig.~\ref{ExpSche} Panel (a). The TLS (qubit) used in our study is encoded in the hyperfine energy levels of the ion, represented as ${\ket{0} \equiv \,^{2}S_{1/2}\, \ket{F=0,m_{F}=0}}$ and ${\ket{1} \equiv \,^{2}S_{1/2}\, \ket{F=1,m_{F}=0}}$, where $F$ denotes the total angular momentum of the atom and $m_{F}$ is its projection along the quantization axis. 
In absence of an external field, the subspace $F=1$ is degenerate. Therefore, we apply an external static magnetic field $\vec{B}$ with intensity $6.40$ G to obtain a $8.9$ MHz Zeeman structure splitting.  This leads to the the desired TLS  with a transition frequency given by $\omega_{hf} = 2\pi \times 12.642~825 $ GHz, see  Fig.~\ref{ExpSche} Panel (b).

The TLS is controlled by a  preprogrammed microwave, which is generated by mixing a  $2\pi \times 12.442$~ GHz coherent local oscillator microwave and a programmable Arbitrary Waveform Generator (AWG) signal, centered around $2\pi \times 200$ MHz~\cite{Hu:18,Hu-18-b}.
This enables implementing the components $\hat{\sigma}_{z}$ and $\hat{\sigma}_{x}$ of the Hamiltonian in Eq. (\ref{eq:Ham}), by simultaneous control of the microwave amplitude and the detuning between microwave frequency $\omega_{\mathbf{0}}$ and the transition frequency $\omega_{\mathbf{hf}}$~\cite{Hu:18}. Here, the Rabi frequency $ \varepsilon\b t$ is directly proportional to the microwave amplitude, and $\omega_{\mathbf{0}} - \omega_{\mathbf{hf}} = \omega(t)$.

To initialize the experiment, first the motion of the ion is cooled by employing a $369.5$~nm Doppler cooling laser beam, using the optical transition cycle $^{2}S_{1/2}\!\rightleftarrows\!^{2}P_{1/2}$. 
During the transition cycle, there is a branching ratio $R$ for population decay from $^{2}P_{1/2}$ state to the $^{2}D_{3/2}$~\cite{Olmschenk:07}. To send the system back to the cooling cycle, a light at $935.2$~nm is used to promote transitions $^{2}D_{3/2}\!\rightleftarrows\!^{3}D[3/2]_{1/2}$, where the system can quickly decay from $^{3}D[3/2]_{1/2}$ to $^{2}S_{1/2}$ (grey arrows in Fig. ~\ref{ExpSche} Panel (b)). 
After Doppler cooling, the system is initialized in the $\ket {0}$ state with a standard optical pumping process. Utilizing the AWG, the time-frequency protocols  of the inertial solutions are implemented.

The measurement procedure detects the population of the excited state of the qubit, using a fluorescence detection, induced by the $369.5$~nm laser  \cite{Hu:18,Hu-18-b}.  Thus, detection of photons correspond to population in the bright state $\ket{1}$, while no photons signify population in the dark state $\ket{0}$, as shown in Fig. \ref{ExpSche},. The overall measurement fidelity is estimated to be $99.4\%$ 
\cite{Hu:18,hu2019quantum}. This experiment is repeated many times, for different delay times and different inertial protocols.   For  each experimental protocol, the normalized energy as a function of time is evaluated $\mean{\hat{H}\b t}/\mean{\hat{H}\b 0}$.

\section{Results}
 
\begin{figure}[t!]
	\centering
	\subfloat{\includegraphics[width=0.35\textwidth]{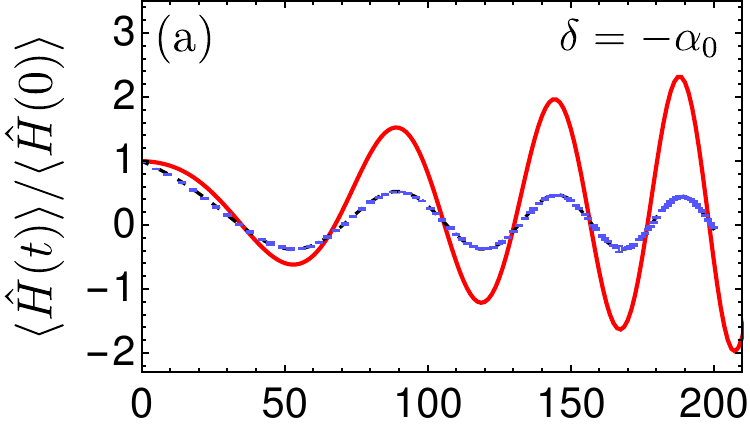}}\quad
	\subfloat{\includegraphics[width=0.35\textwidth]{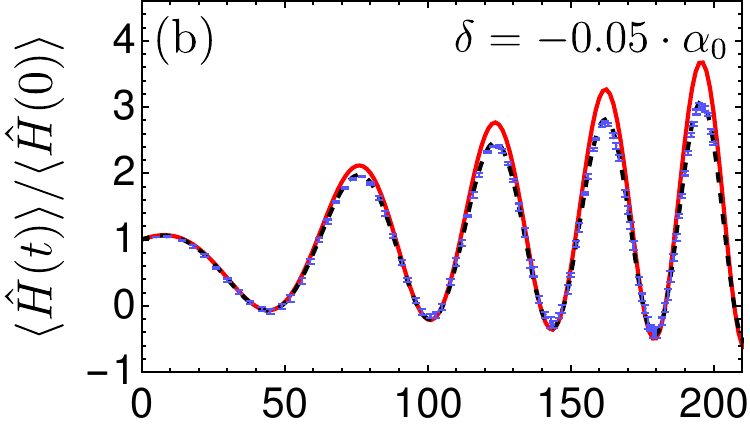}}\quad
	\subfloat{\includegraphics[width=0.35\textwidth]{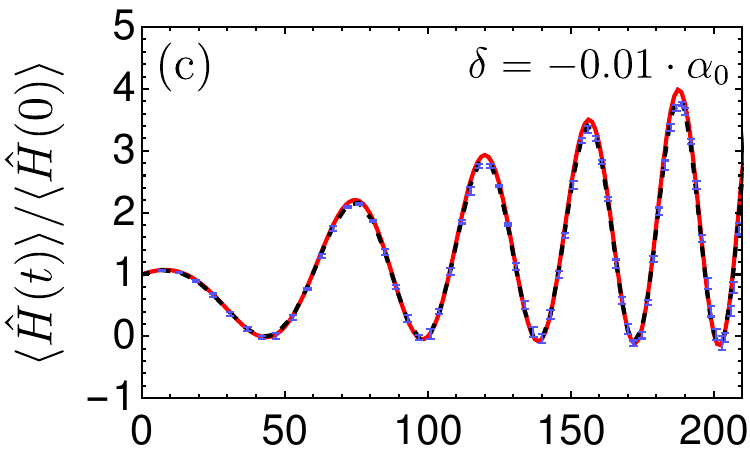}}\quad
	\subfloat{\includegraphics[width=0.35\textwidth]{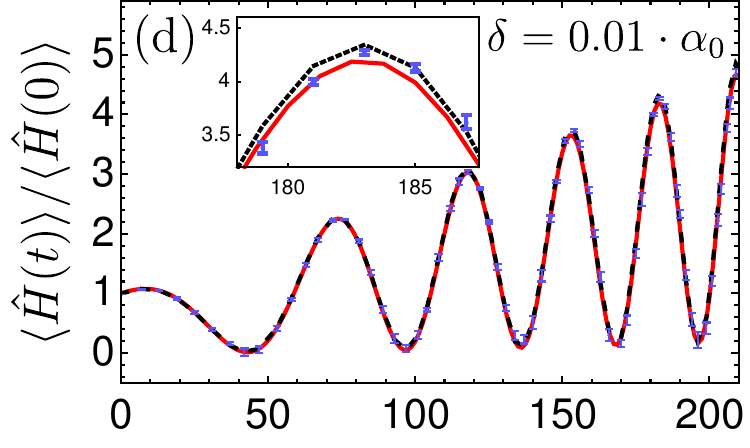}}\quad
	\subfloat{\includegraphics[width=0.35\textwidth]{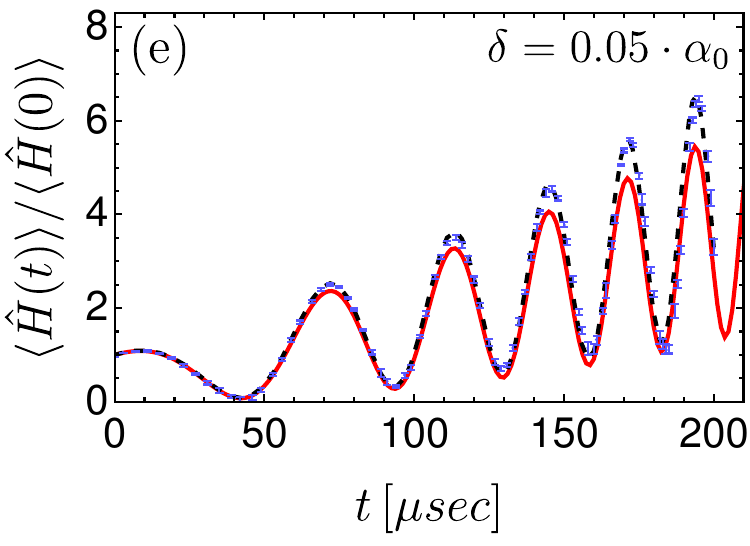}}\quad
	\subfloat{\includegraphics[width=0.35\textwidth]{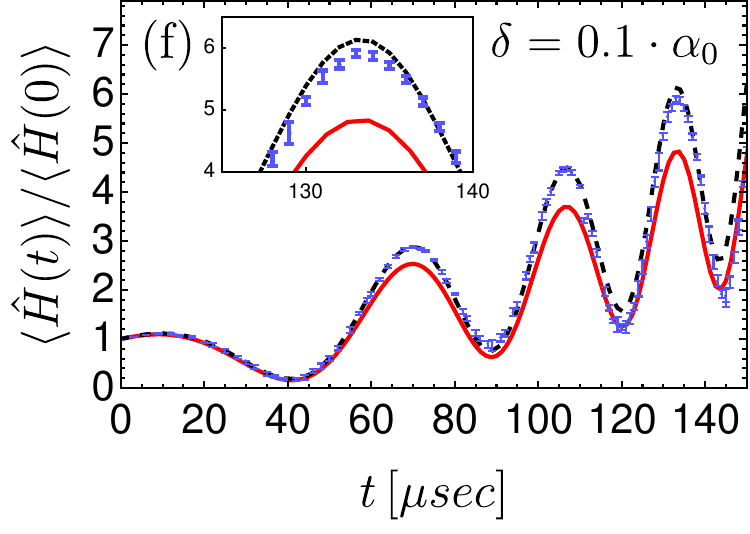}}\quad
	\caption{The normalized energy as a function of time for the experimental result (blue), inertial solution (red) and numerical solution (dashed-black) for different values of $\delta$: (a) $\delta=- \alpha\b{0}$, (b) $\delta=-0.05\cdot\alpha\b{0}$, (c) $\delta=-0.01\cdot\alpha\b{0}$, (d) $\delta=0.01\cdot\alpha\b{0}$, (e) $\delta=0.05\cdot\alpha\b{0}$, (f) $\delta=0.1\cdot\alpha\b{0}$. The experimental parameters are: $\alpha\b 0=6\cdot2\pi KHz$, $\gamma= 50\cdot2\pi M(Hz)^2$ with $\mu\b 0=-1$. The varying values of $|\delta|=|d\mu/dt|$ are related to the quality of the inertial approximation; for slow change in $\mu$, the inertial approximation is satisfied (panels (c) and (d)). Varying $\mu$ rapidly leads to the breakdown of the inertial theorem (see Panels (a),(b),(e) and (f)). The insets in Panel (d) and (f) represent an enlarged section of the last oscillation, highlighting the experimental error bars.} 
	\label{fig:1}
\end{figure}

The qubit's normalized energy as a function of time is shown in Figure \ref{fig:1}, comparing the experimental measurements (blue) to the analytical inertial solution (red) and an exact numerical simulation (black). Experiments with different $\delta$ (Eq. (\ref{eq:protocol_exp})) were realized to asses the range of validity of the inertial solution. There is a good agreement between the theoretical and experimental results for small $\delta$ (see Panel (c) and (d)), demonstrating the high accuracy of the inertial solution. 
When $|\delta|=|d\mu/dt|$ is increased,  we witness the breakdown of the inertial solution (Panels (a),(b),(e) and (f)), the deviations between the predicted normalized energy values of the inertial solution and the experimental results increase. The deviation is manifested by a difference in amplitude, while the phase of the inertial solution follows the exact simulation and experiment measurements, see Sec. \ref{ap:diviation} for a detailed analysis.

\begin{figure}[t!]
\centering
\includegraphics[width=0.7\textwidth]{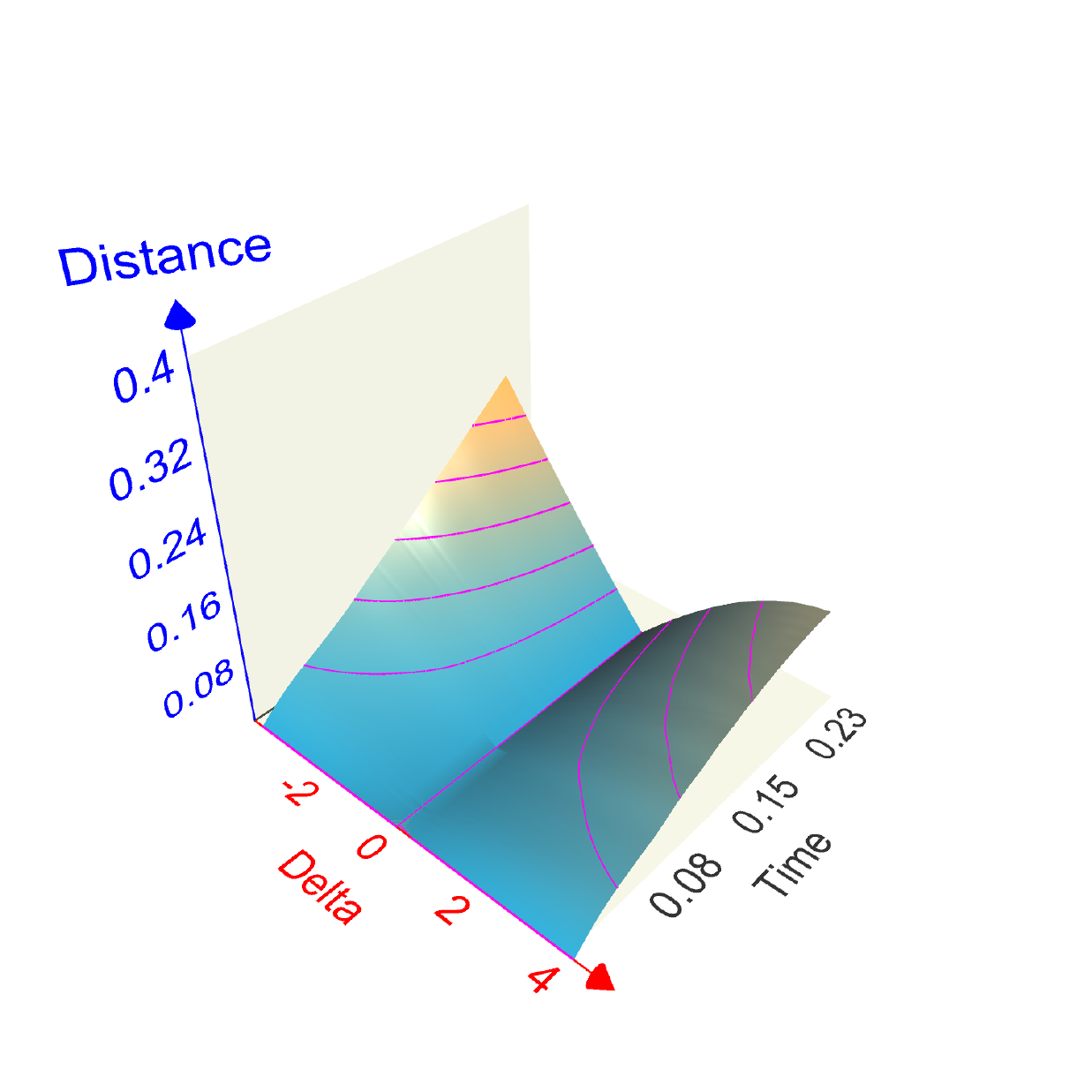}
\caption{Distance ${\cal{D}}$ between the inertial solution and the exact numerical solution as a function of $\delta$ and time. For $\delta=0$, the inertial solution is exact at all times. For larger $|\delta|$, the distance increases almost linearly with time and $|\delta|$. } 
\label{fig:Fid}
\end{figure}

\begin{figure}[t!]
\centering
\includegraphics[width=0.33\textwidth]{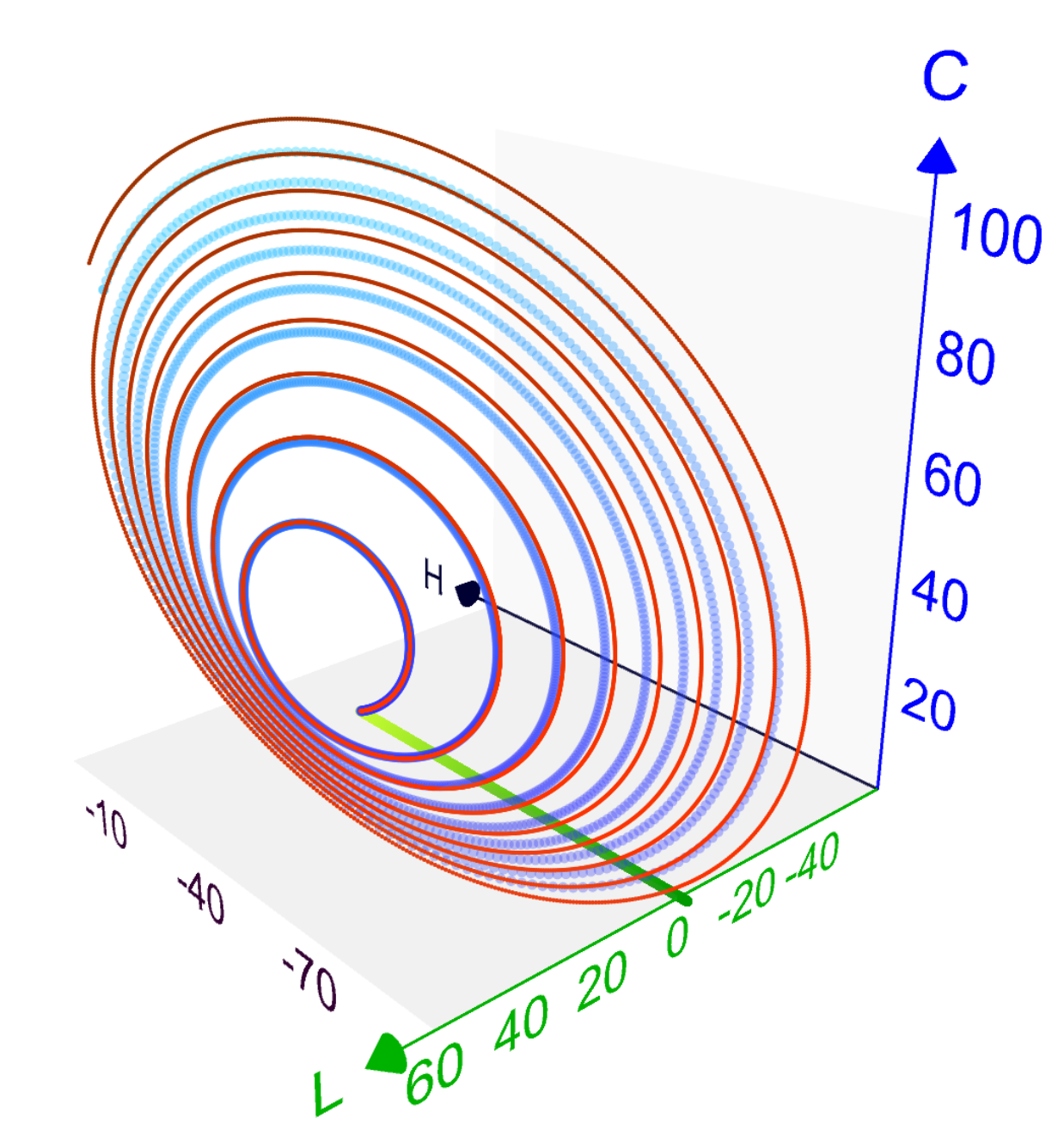}
\includegraphics[width=0.33\textwidth]{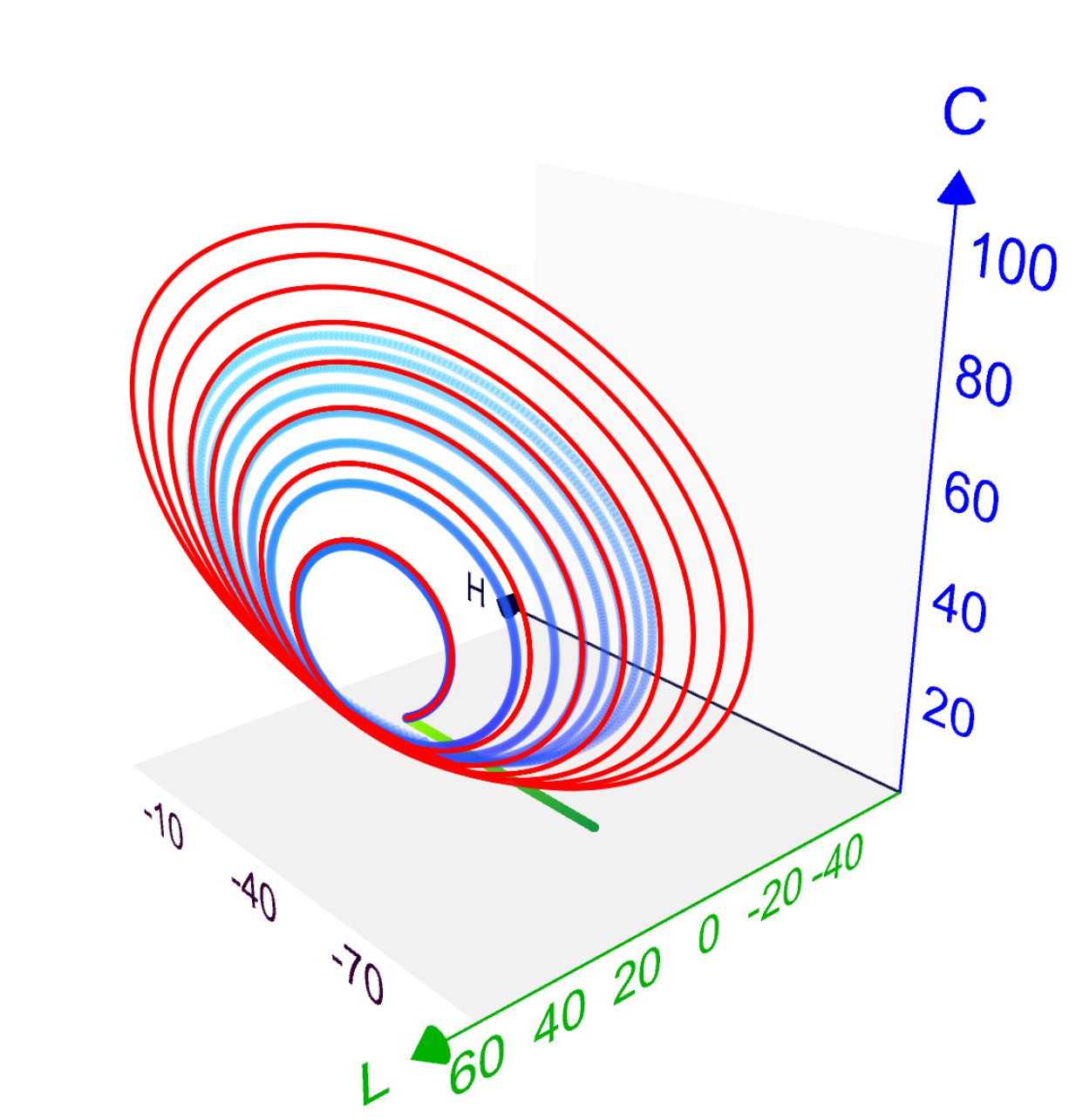}
\caption{The inertial  trajectory (red), exact numerical  (blue) and adiabatic (green straight line) solutions in the $\mean{\hat{H}},\mean{\hat{L}},\mean{\hat{C}}$ coordinate space, 
for (a) $\delta=-0.01\cdot \alpha\b 0$ and (b) $\delta=-0.05\cdot \alpha \b 0$. } 
\label{fig:3D}
\end{figure}

Figure \ref{fig:Fid} shows the distance $\cal{D}$ between the inertial solution and the exact numerical result as a function of $\delta$ and time. $\cal{D}$ is defined as the Euclidean distance between the expectation values of the Liouville state vectors
\begin{equation}
 {\cal{D}}\b t=\sqrt{\sum_{i=1}^3\b{\mean{v^i\b t}-\mean{v_{num}^i\b t}^2}}~~,
\end{equation}
where $v_i$ and $v_{num}^i$ are the $i$'th component of $\v v$ (the inertial solution) and $\v v_{num}$ (the exact numerical solution). When $\mu$ varies slowly, ($\delta=-0.01$) the inertial solution remains exact, whereas for larger absolute values, the numerical and inertial solutions deviate
linearly in $\delta$ and time.
In Fig. \ref{fig:3D}, we present the inertial, numerical and adiabatic trajectories for $\delta=-0.01,-0.05$ in the $\mean{\hat{H}},\mean{\hat{L}},\mean{\hat{C}}$ space. This representation provides a complete description of the dynamics, demonstrating the large deviation between the adiabatic and inertial solutions. 

 \subsection{Deviations from the exact solution}
 \label{ap:diviation}
There are two major sources of deviation between the inertial solution and experimental results. The first is associated with the breakdown of the inertial solution and the second source concerns the inevitable experimental noise. Observing Fig. \ref{fig:1} we find that the major deviation between the  theoretical and experimental results 
is in the amplitude of the energy oscillations, 
while the phase is not affected even for large $|\delta|$ (see for example Panel (a) with $\delta =-\alpha\b 0$). The amplitude of the inertial solution is determined by the real part of the eigenvalues of the propagator. These are dominated by the the general scaling associated with the change in the generalized Rabi frequency, see Eq. (\ref{ap:v u eq}). The imaginary part of the eigenvalues determine the phase.

In order to rationalize the observed deviation we first analyze the correction terms to the inertial solution. 
Gathering Eqs. (\ref{Schro Liouv1}), (\ref{eq:factorized_TLS}), and (\ref{ap:v u eq}) we obtain
\begin{equation}
    \f{d \v u^H \b{\theta}}{d\theta} =-i {\cal{B}}'\b{\mu \b{\theta}} \v u^H \b{\theta}~~.
\label{eq: u theta}
\end{equation}
Next, we define the instantaneous diagonalizing matrix of  ${\cal{B}}'\b{\mu}$, satisfying ${\cal{P}}^{-1}\b{\mu}{\cal{B}}'\b{\mu}{\cal{P}}\b{\mu}={\cal{D}}\b{\mu}$ and the vector $\v{w}^H\b{\theta}={\cal{P}}^{-1}\b{\mu}\v u^H\b{\theta}$.
The dynamics of  $\v w^H\b{\theta}$ are given by 
\begin{equation}
\f{d\v{w}\b{\theta}}{d\theta} =-i{\cal{D}}\v{w} \b{\theta}+{\cal{O}}\v{w} \b{\theta}~~,
    \label{ap:w}
\end{equation}
where ${\cal{O}}=-{\cal{P}}^{-1}\f{{\cal{P}}}{d\theta}$.
For the studied model the diagonalizing matrix of ${\cal{B}}'\b{\mu}$, Eq. (\ref{eq:Bmodel}), obtains the form
\begin{equation}
    {\cal{P}}=\f{1}{2\kappa^2}\left(\begin{array}{ccc}
\frac{1}{\mu} & -\mu & -\mu\\
0 & i\kappa & -i\kappa\\
1 & 1 & 1
\end{array}\right)
\end{equation}
For a slow change in $\mu$, ${\cal{B}}'$ and consequently ${\cal{P}}$ vary slowly with respect to $\theta$. This property allows neglecting the second term in Eq. (\ref{ap:w}), which is qualitatively similar to the inertial approximation.
The deviations from the exact solution are reflected by the term 
${\cal{O}}\b{\theta}={\cal{P}}^{-1}\f{{\cal{P}}}{d\theta}$. Utilizing the identity $\f{d{\cal{P}}}{d\theta}=\f{1}{\Omega}\f{d{\cal{P}}}{d t}$ we obtain
\begin{equation}
    {\cal{O}}= \f{2\mu}{1+\mu^2}\f{d\mu}{d\theta}{\cal{I}}+{\cal{S}}~~,
    \label{eq:O}
\end{equation}
where 
\begin{equation}
   {\cal{S}}= \f{\delta}{2\Omega\kappa^{2}}\left(\begin{array}{ccc}
\frac{1}{\mu} & \mu & \mu\\
-\frac{1}{2\mu} & -\mu & 0\\
-\frac{1}{2\mu} & 0 & -\mu
\end{array}\right)~~.
\end{equation}
Solving the dynamics explicitly leads to
\begin{equation}
    \v{w}\b{\theta}=e^{\b{-i{\cal{D}+\cal{O}}}\theta}\v{w}\b{0}~~~.
\end{equation}
Next, we utilize the  Zassenhaus formula \cite{suzuki1976generalized} to obtain a solution up to first order in $\theta$
\begin{equation}
\v{w}\b{\theta}\approx e^{-i{\cal{D}}\theta}e^{{\cal{O}}\theta}\v{w}\b{0}~~~.
\end{equation}
The correction term to the inertial solution has real eigenvalues, and therefore only influences the amplitude and not the phase. Thus,
the phase  of the inertial solution is not affected even when 
$|d\mu/dt|=|\delta|$ is large.   

The second source of error is a consequence of experimental noise.
We model this noise by a $\delta$-correlated noise in the timing of the driving \cite{PhysRevA.18.1490}.
Such a process is equivalent to adding random noise to the Generalized Rabi frequency $\Omega\b t$, Eq. (\ref{eq:protocol}). 
In the presence of such a noise the effective equation of motion includes  double commutator  in the operator generating the noise \cite{gorini1976completely}.
For timing noise this becomes \cite{kosloff2010optimal}:
\begin{equation}
         \f{d}{dt}\v v^H \b t = -\sb{{i {\cal{M}}\b t} + \Gamma_n^2 {\cal{M}}^2\b t}\v v^H \b t~~,
     \label{noise E of M}
\end{equation}
where the double commutator is represented by ${\cal M}^2$, and $\Gamma_n$ is proportional to the noise amplitude. In this case, the noise has no effect on the eigenoperators with vanishing eigenvalues, $\hat{F}_1$ Eq. (\ref{f-vec})
(the time-dependent constants of motion). The other two eignvalues of the noise ${\cal{M}}^2\b t$ are real and therefore will only influence the amplitude of the signal. 
The experimental results shown  in Fig. \ref{fig:1} 
in particular the insert of Panel (d) and (f)  corroborate this analysis.

\section{Discussion}

The purpose of this study was to establish experimentally a new family of inertial control protocols.
These protocols are experimentally verified using a platform consisting of the hyperfine levels of an Ytterbium ion $^{171}$Yb$^+$ in a Paul trap. This experimental platform is well suited for the evaluation due to its  high fidelity. The high fidelity of both the control field and measurement allow direct comparison with the
theoretical predictions. The inertial theorem provides a family of non-adiabatic protocols that bridge the gap between the sudden and adiabatic limits \cite{dann2021inertial}. Specifically, we studied  control of the $SU(2)$ Lie algebra, which constitutes the single qubit operations. We chose a protocol involving a  chirp in frequency and change in the generalized Rabi frequency, associated with a linear change in the adiabatic parameter $\mu$.

The experiments verify the theorem and the ability to perform inertial protocols. Moreover, as all experiments are influenced by various kinds of noise \cite{childs2001robustness}, the achieved  accuracy confirms the robustness of the inertial solution.
This conclusion is supported by theoretical simulations which verify that the solution is stable to small deviations and noise.

For a larger deviation from the inertial condition ($d\v{\chi}/dt\ra1$) (Fig. \ref{fig:1} panels (a), (b), (e) and (f)), the error first appears in the amplitude, while the phase of the inertial solution is still accurate. 
We confirm this by analyzing a correction to the inertial solution.  
In the $SU(2)$ algebra, the first-order correction in $\theta$ to the phase vanishes (see the discussion following Eq. \ref{eq:O}). 
Incorporating the amplitude correction  into the inertial solution can lead to higher accuracy. The phase information can be utilized for quantum
parameter estimation \cite{helstrom1976quantum} beyond the inertial limit. 

Experimental validation of the inertial solution paves the way to rapid high-precision control. This control can be extended to {\em inertially} driven open systems \cite{dann2021inertial}, utilizing the non-adiabatic master equation \cite{dann2018time}. Such control can regulate the system entropy \cite{dann2018shortcut,dann2020quantum}. 


The present study constitutes a basic step in adding inertial control protocols to the family of constructive mechanisms of control. 
The experimental validation means that inertial protocols cross the barrier between a theoretical entity to laboratory use.
Control based on the inertial theorem can be utilized in rapid applications of quantum information processing  \cite{aharonov2008adiabatic,farhi2000quantum,farhi2001quantum,childs2001robustness} and sensing \cite{perdomo2015quantum}.
\\
 \ack
 We thank KITP for their hospitality. This research was supported by the Adams Fellowship  Program of the Israel Academy of Sciences and Humanities, the National Science Foundation under Grant No. NSF PHY-1748958 and the Israel Science Foundation Grant No.  2244/14, the National Key Research and Development Program of China (No. 2017YFA0304100), National Natural Science Foundation of China (Nos. 61327901, 61490711, 11774335, 11734015), the China Postdoctoral Science Foundation (Grant No. 2020M671861), Anhui Initiative in Quantum Information Technologies (AHY070000, AHY020100), Anhui Provincial Natural Science Foundation (No. 1608085QA22), Key Research Program of Frontier Sciences, CAS (No. QYZDY-SSWSLH003), and the Fundamental Research Funds for the Central Universities (WK2470000026). A.C.S. is supported by S\~ao Paulo Research Foundation (FAPESP) (Grant No 2019/22685-1). A.C.S. acknowledges the partial financial support by the Coordena\c{c}\~ao de Aperfei\c{c}oamento de Pessoal de N\'{\i}vel Superior and the Brazilian National Institute for Science and Technology of Quantum Information (INCT-IQ).

\section*{References}

\bibliographystyle{unsrt}

\end{document}